\documentstyle[12pt]{article}
\title{\baselineskip=9mm
Effects of mass renormalization on the surface properties of
heavy-ion fusion potential}
\author{K. Hagino and N. Takigawa
\\ \\
\medskip
{\it Department of Physics,
Tohoku University, 980--77 Sendai, Japan}\\
}
\date{\today}

\begin{document}
\baselineskip=9mm
\maketitle

\begin{center}
{\bf Abstract}
\end{center}

We discuss the effects of fast nuclear excitations on heavy-ion fusion
reactions at energies near and below the Coulomb barrier.
Using the fusion of two $^{40}$Ca nuclei as an example and
the inversion method, we show that the mass renormalization
induced by fast nuclear excitations
leads to a large surface diffuseness in the effective potential for
heavy-ion fusion reactions.
\medskip

\noindent
PACS number(s):
03.65.Sq, 25.70.Jj, 74.50.+r, 21.10.Re

\medskip

\newpage

Heavy-ion fusion reactions at energies below the Coulomb barrier are
typical examples of the macroscopic quantum tunneling,
which has been a very popular subject in the past decade
in many fields of physics and chemistry\cite{CL81,HTB90,JJAP93}.
One of the major interests in the macroscopic
quantum tunneling is to assess the effects of environments on the
tunneling rate of a macroscopic variable.
When the bombarding energy is below the Coulomb barrier, heavy-ion
fusion reactions take place by a quantum tunneling.
It is now well established that the fusion cross section at sub-barrier
energies is enhanced by several orders of magnitude compared with
the prediction of a one-dimensional potential model
due to the coupling of the relative motion to nuclear
internal degrees of freedom, which play the role of environments\cite{B88}.
The present authors have shown that the coupling of the tunneling
degree of freedom to fast environmental degrees of freedom
causes both a static potential and a mass renormalizations\cite{THAB94}.
On the other hand, the first and the
second derivatives of the experimental fusion cross section
with respect to the bombarding energy suggests
that the fusion reaction between two $^{40}$Ca nuclei
can be described by a potential model using the surface diffuseness
parameter $a$=1.27 fm\cite{RD93,RKL89}, which is much larger than the
usually accepted value $a\sim$ 0.6 fm.
The spin distribution of the compound nucleus made by the fusion
of $^{16}$O with $^{154}$Sm
also suggests a large effective surface
diffuseness parameter $a$=1.27 fm\cite{RLWL93}.
Such a large surface diffuseness has already been
suggested long time ago through the inversion method
of obtaining the effective fusion potential directly from
the experimental data. In that procedure,
a much larger surface diffuseness $a$=1.59 fm was suggested
for $^{40}$Ca + $^{40}$Ca scattering \cite{BKN83}.

The aim of this paper is to discuss the connection between this large
effective surface diffuseness and the potential and the mass renormalizations
induced by the coupling of the relative motion between heavy ions with fast
nuclear intrinsic degrees of freedom.
We consider the fusion of two $^{40}$Ca nuclei as an example.
We first determine the bare nuclear
potential in the entrance channel by the double folding procedure.
We then calculate the excitation function of the fusion cross section
by taking the effects of a few vibrational modes of excitation
of $^{40}$Ca into account through the adiabatic potential
as well as the mass renormalizations.
We then use the inversion procedure to obtain the effective fusion potential
from thus obtained excitation function of the fusion cross section
and discuss its surface properties.

In the double folding procedure, one often
expresses the densities of nuclei by a step function of range $R_d$
convoluted with a Yukawa smearing function of range $1/\kappa_d$ and
assume a Yukawa interaction for the nucleon-nucleon interaction
with the range parameter $\kappa_I$.
The resulting potential consists of two parts, which are
proportional to $e^{-\kappa_d R}$ and $e^{-\kappa_I R}$
in the surface region\cite{BW81}. If the Michigan three range
Yukawa (M3Y) interaction\cite{SL79,BBML77} is assumed
as is often done for the nucleon-nucleon interaction,
$\kappa_I$ is larger than the standard value of $\kappa_d$.
The diffuseness parameter of the double folding potential is
then approximately given by the inverse of $\kappa_d$.
More precisely, the surface diffuseness parameter slightly deviates
from this simple consideration.
Actually, Aky\"uz and Winther numerically performed the
double folding procedure and obtained the value 1/1.36 fm for
the surface diffuseness parameter in the potential
between two $^{40}$Ca nuclei when they
assumed the charge
distribution of $^{40}$Ca to be given by  $\kappa_d$=1.55 fm$^{-1}$
and used the M3Y force
as the nucleon-nucleon interaction\cite{AW79}.
Note that 1/1.36=0.75 fm is slightly larger than the standards of what is
usually used in optical-model calculations, but not so much as those
suggested from the analyses of the experimental sub-barrier fusion
cross section. Henceforth we take the Woods-Saxon shape
for the bare nuclear potential between two $^{40}$Ca nuclei
\begin{equation}
V_N(R)=-\frac{V_0}{1+e^{(R-2R_0)/a}}
\end{equation}
and use the value 0.75 fm for the surface diffuseness parameter $a$ as
suggested in the above arguments. Following ref.\cite{BW81},
the strength and the range of the bare nuclear potential are taken to be
\begin{eqnarray}
V_0&=&16\pi\gamma a\bar{R} \\
R_0&=&1.2\cdot 40^{1/3}-0.09,
\end{eqnarray}
where $\gamma$ and $\bar{R}$ are 0.95 and $R_0/2$, respectively.
The bare nuclear potential thus obtained gives
the Colulomb barrier at 10.0fm with the height and the
curviture 53.1 MeV and 3.43 MeV, respectively.

We now consider the effects of surface vibrations
of $^{40}$Ca during the fusion process.
We ignore the effects of nucleon transfer reactions, althogh they play
an important role in quantitatively explaining the experimental
fusion cross section \cite{EFL89,LDBP85,BRBW88}.
Following ref.\cite{EFL89} we consider the excited states of
$^{40}$Ca at 3.74, 3.91, and 4.49 MeV with spin-parity
3$^-$, 2$^+$ and 5$^-$ to be the one phonon state of three
different vibrational modes of excitation, though the 2$^+$ state
will rather be the 2$^+$ member of the rotational band based on the 0$^+$
state at 3.35 MeV\cite{ROM90}.
Our interest is to study the effects of the coupling of the relative
motion to these states on the surface diffuseness parameter in the effective
potential for fusion reactions.

We assume the linear oscillator coupling and
use the no-Colioris approximation in order to reduce the dimension
of the coupled-channels equations\cite{HTBB95,TI86}.
The effective Hamiltonian for this system then reads
\begin{eqnarray}
H&=&-\frac{\hbar^2}{2M}\left(\frac{\partial ^2}{\partial R^2}+\frac{2}{R}
\frac{\partial}{\partial R}\right)
+\frac{J(J+1)\hbar^2}{2MR^2}+V_C(R)+V_N(R) \nonumber \\
&&+\sum_{\lambda=3,2,5}
\left\{\hbar\omega_{\lambda}\left(a_{\lambda}^{\dagger}a_{\lambda}+
\frac{1}{2}\right)
+\sqrt{\frac{2\lambda+1}{4\pi}}
\left(\alpha_{\lambda 0}^{(N)}f_N(R)+\alpha_{\lambda 0}^{(C)}f_C(R)\right)
(a_{\lambda}^{\dagger}+a_{\lambda})\right\}
\end{eqnarray}
where $M, R, J, V_C, V_N,$ and $\lambda$ are the reduced mass, the
relative distance, the total angular momentum, the Coulomb and
the nuclear interactions between the two $^{40}$Ca nuclei, and the
multipolalities of the vibrational excitations, respectively.
$f_N(R)$ and $f_C(R)$ are the nuclear and the Coulomb coupling form
factors, respectively.
We determine them following the collective model\cite{BW81}.
$\hbar\omega_{\lambda}, a_{\lambda}^{\dagger}(a_{\lambda}),
\alpha_{\lambda 0}^{(N)}$, and $\alpha_{\lambda 0}^{(C)}$ are
the excitation energy, the creation (annihilation) operators
of the oscillator quanta, the nuclear and the Coulomb
coupling strengthes of the 2$^{\lambda}$-pole vibrational excitations,
respectively.
The factor $\sqrt{\frac{2\lambda+1}{4\pi}}$ in front of the
coupling strength arises from the no-Coriolis approximation.
We take the coupling strengthes from
ref.\cite{EFL89} (see Table I).
In Table I, $R_C$ and $R_N$ are the charge and the matter
radii of $^{40}$Ca, respectively.
The Coulomb coupling strength was determined from the transition
strength $B(E\lambda)$ from the one phonon state of each
vibrational mode of excitation to the ground state.
The nuclear coupling strength was modified
to reproduce the data of inelastic scattering of $^{16}$O from
$^{40}$Ca. All the coupling strengthes are multipled by $\sqrt{2}$
in order to take the mutual excitations of the target and the projectile
into account.

The excitation energy $\hbar\omega$ of all the vibrational excitations
which we are interested in now is larger
than the curvature of the bare potential barrier. Therefore, the present
problem corresponds to the case of an adiabatic tunnelig,
i.e. a slow tunneling.
The adiabaticity of the tunneling process depends on two
parameters\cite{THA94}.
One is the ratio of the energy scale $a_1=\hbar\Omega / \hbar\omega$,
$\hbar\Omega$ being the curvature of the bare potential barrier.
The other is the ratio of the strength of the coupling Hamiltonian
to the excitation energy $a_2=\alpha_0cR_a/\hbar\omega$, $R_a$ and
$\alpha_0c$ being the thickness of the tunneling region and
the derivative of the coupling Hamiltonian
in the case when one assumes a bi-linear coupling, respectively.
The adiabatic assumption breaks down if these parameters are larger than one.
The typical values of these parameters for $^{40}$Ca + $^{40}$Ca
scattering are listed in Table II.
In order to obtain $a_2$, we estimated the derivative of the
coupling Hamiltonian at the position of the bare
potential barrier and determined $R_a$
at energy 1 MeV below the top of the bare potential barrier
based on the parabolic approximation.
Table II justifies the use of the adiabatic formula to calculate
the barrier penetrability for the fusion reaction in the present system.
We use the revised adiabatic formula of ref.\cite{THAB94}.
It modifies the well known adiabatic formula,
which takes only the static potential shift into account,
by adding the mass renormalization.
Our formula for the barrier penetrability then reads
\begin{equation}
P(E)=\frac{1}{1+e^{2S(E)}}
\end{equation}
with
\begin{equation}
S(E)=\int_{R_1}^{R_2}dR \sqrt{\frac{2M_{ad}(R)}{\hbar^2}\left(V_{ad}(R)
-E\right)}
\end{equation}
for the $s$-wave scattering. In eq.(6) $R_1$ and $R_2$ are the
inner and the outer classical turning points, respectively. The
renormalized potential and mass in this equation
are given by \cite{THAB94}.
\begin{eqnarray}
V_{ad}(R)&=&V_C(R)+V_N(R)-
\sum_{\lambda=3,2,5}
\frac{2\lambda+1}{4\pi}
\left(\alpha_{\lambda 0}^{(N)}f_N(R)+\alpha_{\lambda 0}^{(C)}f_C(R)\right)^2
\frac{1}{\hbar\omega_{\lambda}} \\
M_{ad}(R)&=&M+2
\sum_{\lambda=3,2,5}
\frac{2\lambda+1}{4\pi}
\left(\alpha_{\lambda 0}^{(N)}\frac{df_N(R)}{dR}
+\alpha_{\lambda 0}^{(C)}\frac{df_C(R)}{dR}\right)^2
\frac{1}{\hbar\omega_{\lambda}^3}
\end{eqnarray}

Fig.1 shows the tunneling probability in the $s$-wave scattering
as a function of the bombarding energy.
The dotted line is the penetrability
when there is no coupling, i.e. the penetrability for the
bare potential barrier with the bare mass $M$.
The dashed line is obtained with the standard adiabatic formula,
where only the adiabatic potential renormalization is
taken into account.
The result of the revised adiabatic formula which takes the mass
renormalization into account is denoted by the solid line.
Notice that the overenhancement of the effects of the channel coupling
when one considers only the adiabatic potential shift (the dashed line)
 is corrected by the effects of the mass renormalization.
Fig.2 shows the renormalized mass $M_{ad}(R)$ in this system (eq.(8))
in the ratio to the bare mass $M$.
It is a function of the separation distance
between the colliding nuclei reflecting the radial dependence
of the coupling form factor. It has two
peaks because of the strong cancellation between
the Coulomb and the nuclear couplings.
The small spike at around $R$=8fm originates from the discontinuity of the
derivative of the Coulomb coupling form factor.
This defect plays, however, no essential role.

We now derive the effective fusion potential corresponding to the
solid line in Fig.1. We thus demonstrate how the effects of the channel
coupling including the mass renormalization can be mocked up by
renormalizing the surface diffuseness.
To this end, we use the
inversion procedure which gives the thickness of the tunneling region,
i.e. the thickness of the potential barrier, at each
energy directly from the excitation function of the
fusion cross section\cite{BKN83}
\begin{equation}
t(E)=-\frac{2}{\pi}\sqrt{\frac{\hbar^2}{2M}}
\int^B_E\frac{\frac{dS(E')}{dE'}}{\sqrt{E'-E}}dE'
\end{equation}
where $S(E)$ is the action integral given by eq.(6) and $B$ is the
barrier height detemined from the condition $S(B)=0$.
This is the height of the potential barrier
including the adiabatic potential shift.
Note that the mass $M$ in the factor in front of the integral in eq.(9)
does not include the mass renormalization, but the bare mass.
This is because our aim is to get the effective potential for the
relative motion with bare mass.
The results of the inversion procedure are shown in fig.3 by the
solid line. For the integration in eq.(9), we changed the variable
from $E'$ to $\sqrt{E'-E}$ to eliminate the singularity
at the lower limit of the integral\cite{BKN83}.
The inversion procedure gives only the barrier thickness $t(E)$.
Consequently, the turning points themselves as functions of the energy
remain indeterminate.
We assumed that the outer turning points are given by those of the
adiabatic potential barrier and determine the inner turning points
of the effective potential barrier by using the barrier thickness
given by the inversion procedure.
For comparison, fig.3 also shows the bare potential
barrier (the dotted line) and the adiabatic
potential barrier (the dashed line).

Since we are interested in the surface diffuseness parameter of the
nuclear potential, we subtract the Coulomb potential $V_C(R)$ from
both the effective potential obtained by the inversion procedure and
the adiabatic potential. We call these potentials the
effective nuclear potentials, because they contain not only the effects
of the nuclear coupling but also those of the Coulomb coupling.
Fig.4 shows the effective nuclear potentials obtained in this way.
The dashed line denotes the effective nuclear
potential in the adiabatic limit, i.e., the potential
obtained by subtracting the Coulomb potential from the
adiabatic potential.
The effective nuclear potential, which takes account of
the mass renormaliztion as well, is denoted by the solid line.
Also shown are the potentials of
an exponential shape with two different surface
diffuseness parameters.
Their strength was chosen, such that the values of the resultant
potentials agree with that of the adiabatic potential
at the peak position.
The dot-dashed and the dotted lines correspond to the
potential when the surface diffuseness parameter was chosen to be
0.7 and 0.8 fm, respectively.
These choices simulate the effective potential in the adiabatic limit
(the dashed line), and that which includes the effects of the mass
renormalization(solid line).
Remember that the surface diffuseness parameter for the bare potential
is about 0.75 fm. Fig.4 shows that the surface diffuseness
parameter is smaller than that of the bare nuclear potential if
only the effects of the adiabatic potential shift is taken into account, but
gets larger than that of the
bare nuclear potential when the effects of the
mass renormalization is added.
We thus conclude that one of
the origins of the large surface diffuseness of the effective
nuclear potential in heavy-ion fusion reactions
is the effects of the coupling to nuclear inelastic excitations,
which lead to both the potential and the mass renormalizations.

In summary, we discussed the effects of nuclear intrinsic
excitations on heavy-ion fusion reactions. We recast
the potential and the mass renormalizations
due to fast inelastic excitations
of scattering nuclei in terms of the renormalized surface diffuseness
in the effective fusion potential.
As a concrete example, we analyzed the fusion reactions between
two $^{40}$Ca nuclei.
We used the inversion method to obtain the effective
nuclear potential and found that the
surface diffuseness parameter becomes larger than that
obtained by the double folding method if both the adiabatic
potential shift and the adiabatic mass renormalization are taken into account.
The value of the surface diffuseness parameter we obtained is
about 0.8 fm compared to the standard value 0.6 fm.
Although this value is not sufficiently large  to explain the
large surface diffuseness empirically obtained from the experimental data of
sub-barrier fusion cross sections, we conclude that
the large surface diffuseness obtained from data analyses
partly represents the effects of channel coupling.
It would be interesting to study how the effects of transfer reactions
further modify the surface properties of the effective fusion potential.
The large surface diffuseness parameter has also been suggested
in the fusion reactions of $^{16}$O on $^{154}$Sm, which
is a fast tunneling rather than a slow tunneling discussed
in this paper\cite{RLWL93}.
It remains an open question how we should take into account the
effects of the channel coupling in such a system to
explain the large surface diffuseness.
Contrary to the case of
slow tunneling discussed in this paper, one has to treat
a distributed potential barriers, or an energy dependent
effective potential barrier\cite{HTBB94}.

\medskip

The authors would like to thank N. Rowley, A.B. Balantekin and
S. Yoshida for useful
discussions.
The work of K.H. was supported by Research Fellowships
of the Japan Society for the Promotion of Science for
Young Scientists.
This work was supported by the Grant-in-Aid for General
Scientific Research,
Contract No.06640368, the Grant-in-Aid for Scientific
Research on Priority
Areas,Contract No.05243101,from the Japanese Ministry of Education,
Science and Culture.

\newpage

\begin{center}
Table I
\end{center}

\begin{table}[htb]
\begin{center}
\begin{tabular}{lccc}\hline
{$\lambda^{\pi}$}&
{$\hbar\omega_{\lambda}$(MeV)}&
{$R_C\alpha_{\lambda 0}^{(C)}$(fm)}&
{$R_N\alpha_{\lambda 0}^{(N)}$(fm)} \\ \hline
3$^{-}$ & 3.737 & 0.6576 & 0.4455 \\
2$^{+}$ & 3.905 & 0.1952 & 0.1768 \\
5$^{-}$ & 4.492 & 0.4837 & 0.2475 \\ \hline
\end{tabular}
\end{center}
\end{table}

\begin{center}
Table II
\end{center}

\begin{table}[htb]
\begin{center}
\begin{tabular}{lcc}\hline
{$\lambda^{\pi}$}&{$a_1$}&{$a_2$} \\ \hline
3$^-$ & 0.9177 & 0.6176  \\
2$^+$  & 0.8783 & 0.1997  \\
5$^-$ & 0.7635 & 0.3668  \\ \hline
\end{tabular}
\end{center}
\end{table}

\bigskip

\begin{center}
{\bf Table Captions}
\end{center}

\noindent
{\bf Table I:} Low lying vibrational states of $^{40}$Ca.
$\hbar\omega_{\lambda}$, $\alpha_{\lambda 0}^{(C)}$ and
$\alpha_{\lambda 0}^{(N)}$
are the excitation energy of the 2$^{\lambda}$-pole
vibrational state, and its Coulomb and nuclear coupling strengthes,
 respectively.
$R_C$ and $R_N$ are the charge and the matter radii,
respectively.

\noindent
{\bf Table II:} Adiabaticity parameters for the
tunneling process. $a_1$ is the ratio of the energy scales
of the tunneling and the environmental
degrees of freedom $\hbar\Omega/\hbar\omega$, while
$a_2$ the ratio of the strengthes of
the channel coupling Hamiltonian to the excitation energy
of the environment $\alpha_0cR_a/\hbar\omega$.

\newpage

\begin{center}
{\bf Figure Captions}
\end{center}

\noindent
{\bf Fig.1:} Excitation function of the s-wave barrier penetrability
for $^{40}$Ca + $^{40}$Ca scattering.
The dotted line is the penetrability in the absence of the
channel coupling.
The dashed line was obtained by the standard adiabatic formula which
takes only the effects of the adiabatic potential shift into account.
The solid line was calculated with the revised adiabatic formula
by adding the effects of the mass renormalization.

\noindent
{\bf Fig.2:} The renormalized mass (eq. (8)) as a function of the
separation distance between the colliding nuclei in the
unit of the bare mass.

\noindent
{\bf Fig.3:} Fusion barrier obtained by the inversion procedure.
The dotted and the dashed lines are the bare
and the adiabatic potential barriers, respectively.
The solid line includes the effects of the mass renormalization.

\noindent
{\bf Fig.4:} Effective nuclear potential for the fusion of
two $^{40}$Ca nuclei. The solid line was
obtained by subtracting the Coulomb potential $V_C(R)$ from the
effective potential barrier in the presence of the mass renormalization.
The dashed line is the effective nuclear potential in the adiabatic
limit disregarding the mass renormalization.
The dott-dashed and the dotted lines are the
exponential potentials with the surface
diffuseness parameters 0.7 and 0.8 fm,
respectively. Their depth was chosen to reproduce the barrier height
of the adiabatic potential. The surface diffuseness parameters
$a$=0.7 fm and 0.8 fm were fixed to simulate the nuclear potential
without (the dashed line) and with (the solid line) the mass renormalization.

\end{document}